\documentclass[conference]{IEEEtran}
\IEEEoverridecommandlockouts
\usepackage{cite}
\usepackage{amsmath,amssymb,amsfonts}
\usepackage{algorithmic}
\usepackage{textcomp}
\usepackage{xcolor}
\ifCLASSINFOpdf
  \usepackage[pdftex]{graphicx}
  \graphicspath{{figures/}}
  \DeclareGraphicsExtensions{.eps,.jpeg,.png,.pdf}
\else
  \usepackage[dvips]{graphicx}
  \graphicspath{{figures/}}
  \DeclareGraphicsExtensions{.eps,.jpeg,.png,.pdf}
\fi

\usepackage{tikz}
\usepackage{standalone}
\usepackage{amsmath, amsfonts, bm, amsthm, amssymb}
\usepackage{algorithmic}
\usepackage[ruled,vlined,linesnumbered]{algorithm2e}

\usepackage{array}
\usepackage{blkarray}
\usepackage{tabularx,booktabs}

\ifCLASSOPTIONcompsoc
 \usepackage[caption=false,font=normalsize,labelfont=sf,textfont=sf]{subfig}
\else
 \usepackage[caption=false,font=footnotesize]{subfig}
\fi

\usepackage{dblfloatfix}

\usepackage{url}

\usepackage{enumitem}

\usepackage{flushend}

\newlist{inlineroman}{enumerate*}{1}
\setlist[inlineroman]{afterlabel=~,label=\roman*)}

\newtheorem{thm}{Theorem}

\newcommand{\mch}[2]{
\left.\mathchoice
  {\left(\kern-0.48em\binom{#1}{#2}\kern-0.48em\right)}
  {\big(\kern-0.30em\binom{\smash{#1}}{\smash{#2}}\kern-0.30em\big)}
  {\left(\kern-0.30em\binom{\smash{#1}}{\smash{#2}}\kern-0.30em\right)}
  {\left(\kern-0.30em\binom{\smash{#1}}{\smash{#2}}\kern-0.30em\right)}
\right.}

\def\BibTeX{{\rm B\kern-.05em{\sc i\kern-.025em b}\kern-.08em
    T\kern-.1667em\lower.7ex\hbox{E}\kern-.125emX}}
\begin{document}

\title{
Graph Neural Networks for Physical-Layer Security in Multi-User Flexible-Duplex  Networks

 \vspace{-0.3cm}}

\author{\IEEEauthorblockN{Tharaka Perera\IEEEauthorrefmark{1},
Saman Atapattu\IEEEauthorrefmark{2}, Yuting Fang\IEEEauthorrefmark{1}, and
Jamie Evans\IEEEauthorrefmark{1}}%
\IEEEauthorblockA{\IEEEauthorrefmark{1}Department of Electrical and Electronic Engineering, University of Melbourne, Victoria, Australia. \\
\IEEEauthorrefmark{2}School of Engineering, RMIT University, Melbourne, Victoria, Australia. \\
Email: \IEEEauthorrefmark{1}\{perera.t, yuting.fang, jse\}@unimelb.edu.au,
\IEEEauthorrefmark{2}saman.atapattu@rmit.edu.au
}
}

\maketitle

\begin{abstract}

This paper explores Physical-Layer Security (PLS) in Flexible Duplex (FlexD) networks, considering scenarios involving eavesdroppers. Our investigation revolves around the intricacies of the sum secrecy rate maximization problem, particularly when faced with coordinated and distributed eavesdroppers employing a Minimum Mean Square Error (MMSE) receiver. Our contributions include an iterative classical optimization solution and an unsupervised learning strategy based on Graph Neural Networks (GNNs). To the best of our knowledge, this work marks the initial exploration of GNNs for PLS applications. Additionally, we extend the GNN approach to address the absence of eavesdroppers' channel knowledge. Extensive numerical simulations highlight FlexD's superiority over Half-Duplex (HD) communications and the GNN approach's superiority over the classical method in both performance and time complexity.

\end{abstract}

\vspace{-0.2em}
\section{Introduction}
Flexible Duplex (FlexD) networks improve spectral efficiency compared to traditional half-duplex (HD) and full-duplex (FD) systems by dynamically adapting time and frequency allocations. Unlike fixed resource blocks for uplink (UL) and downlink (DL) in HD communications, FlexD dynamically adjusts to channel changes. Recent advancements like resource allocation and game-theoretic-mode scheduling enhance network throughput \cite{Dayarathna2021, flex_net_2022, d_tdd_game_2021, phy_cell_2021}. In the context of robust wireless connectivity beyond 5G, emphasizing security is crucial \cite{Mucchi2021ojcs}. Our research focuses on PLS with FlexD strategy to address evolving security challenges in wireless applications.

\subsection{Related work}
Complex wireless networks pose a challenge in securing data against intelligent eavesdroppers. Traditional cryptography is often insufficient. Physical layer security (PLS) leverages the dynamic wireless channels to establish secure connections, operating independently of upper-layer integrations. This simplifies network architectures and enhances security \cite{atapattu2019source,wen2023covert, ari2022performance}.  PLS has been extensively investigated for both conventional HD and FD communications, as discussed in \cite{zheng2017physical,atapattu2019physical,nuradha2019physical,wijewardena2021physical} and references therein.
The literature on FlexD presents diverse algorithms for resource scheduling, including a graph neural network (GNN) approach \cite{flex_net_2022}, iterative pattern search algorithms \cite{Dayarathna2021}, a radio frame selection algorithm for flexible duplex networks \cite{liu2015performance}, and resource management in Narrowband Internet of Things (NB-IoT) \cite{malik2019radio}. However, the dynamic scheduling demands in FlexD networks pose a computationally complex challenge due to the combinatorial nature, exacerbated by the need for secure communication in the presence of eavesdroppers. Notably, none of the existing approaches specifically address the goal of PLS-targeted resource allocation in FlexD networks.

\subsection{Problem statement and contribution}
FlexD-based PLS in multi-user networks necessitates effective power allocation and communication direction. The non-convex nature of power allocation and the combinatorial challenges in dynamic scheduling make classical algorithms less efficient. Recent advancements in wireless communication leverage machine learning to tackle such issues in complex networks, extending to resource allocation \cite{guan2021customized}, wireless sensing \cite{wang2020learning}, and user localization \cite{zhang2021indoor}. The intrinsic graph structure of wireless networks aligns with the use of GNNs, tailored for processing non-Euclidean data. Recent GNN applications encompass decentralized inference \cite{lee2021decentralized}, resource allocation with random edge GNNs \cite{eisen2020optimal}, and link scheduling \cite{zhao2022link}. 

In addressing the research gap between multi-user FlexD networks and PLS, this work pioneers the introduction of a novel multi-user FlexD system model. \textit{It further proposes both classical optimization and GNN algorithms as the initial endeavors to maximize network secrecy within the existing literature.} This study presents several noteworthy contributions:
\begin{enumerate}
    \item {\it Innovative System Model:} Presenting a novel system model that accommodates multiple legitimate users and coordinated eavesdroppers. Assuming optimal conditions for eavesdroppers, we set a lower bound for the system's sum secrecy rate, initially addressing the optimization problem with classical methods.
    \item {\it Efficient Graph Representation of FlexD System:} Introducing a distinctive graph representation for the FlexD system, consolidating two paired users into a single node to optimize memory and computational resources. Then, integrating passive eavesdropper nodes into user pairs reduces node count and simplifies the network graph.
    \item {\it GNN-Based Optimization Strategy:} Introducing an innovative GNN-based optimization to maximize the sum secrecy rate of FlexD networks. This approach outperforms classical methods in performance and time complexity.
    \item {\it Enhanced GNN Model:} Modifying the GNN model to predict without eavesdropper channel information achieves comparable performance to models with full channel data, enhancing practicality and versatility.
\end{enumerate}

\noindent\textbf{Notation.} 
Scalars: lowercase, Vectors: bold lowercase, Matrices: bold uppercase. Conjugate transpose: $(\cdot)^\dag$, Identity matrix: $\bm{I}_V$ ($V \times V$). Sets: $\{\cdot\}$, Multisets: $\{\!\!\{\cdot\}\!\!\}$, Vector concatenation:~$|\!|$.

\section{System Model}

\begin{figure}[!t]
    \centering
    \includegraphics[width=0.75\linewidth]{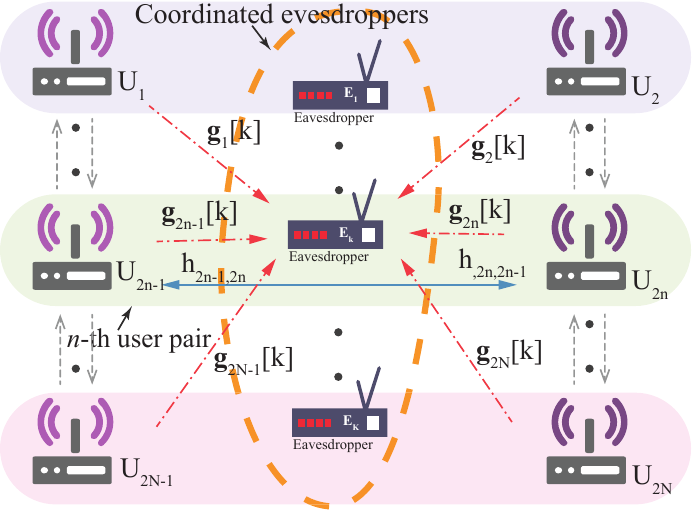}
    \caption{FlexD network with $2N$ users and $K$ coordinated eavesdroppers.}
    \label{fig:system_model_mul}
    \vspace{-1.2em}
\end{figure}
\subsection{Network Model}
As illustrated in Fig.~\ref{fig:system_model_mul}, the FlexD network comprises $2N$ users, denoted as $U_1, \dots, U_{2N}$, aiming for simultaneous secure communications amid the presence of $K$ coordinating eavesdroppers, labeled as $E_1, \dots, E_K$. In Fig.~\ref{fig:system_model_mul}, solid lines represent desired communication links, dashed lines indicate interference links, and dash-dotted lines denote leakage links dedicated to eavesdropping. For clarity, the figure specifically highlights channels involving $U_{2n-1}$, $U_n$, and  $E_k$.
Within the FlexD network, without loss of generality, users with adjacent indices form user pairs, denoted as $(1\leftrightarrow2, \dotsc, (2n-1)\leftrightarrow2n, \dotsc, (2N-1)\leftrightarrow2N)$. In a resource block, each user pair designates one node as the transmitter and the other as the receiver. FlexD users managing interference as receivers face $(N-1)$ transmitter nodes. As transmitters, they minimize leakage to $K$ eavesdroppers and interference to other users. FlexD user pairs dynamically schedule communication directions, capitalizing on wireless channel variations to maximize sum secrecy rates.

Eavesdroppers within the network engage in passive monitoring of FlexD user pair communications.  Our analysis assumes adaptive coordinated eavesdroppers, strategically adjusting their positions to optimize the signal-to-interference plus noise ratio (SINR), thereby maximizing the potential for secure communication leakage. Eavesdroppers employ a Minimum Mean Square Error (MMSE) receiver, treating interference as noise, representing the best-case scenario with Gaussian codebook sequences \cite{bustin2012mmse}. For simplicity, we assume single-antenna configurations for FlexD users and eavesdroppers, though the analysis can extend to multi-antenna scenarios.

\vspace{-0.3em}
\subsection{Analytical Model}
Consider a FlexD user pair denoted as $U_m-U_n$, where $n\in{1,\cdots,2N}$ and $m = 2(n\bmod2) + n - 1$. For a given resource block, the SINR of the $n$th FlexD node is expressed as 
\vspace{-0.3em}
\begin{equation}
    \gamma_n^F = \frac{t_m p_m |h_{nm}|^2}{\sigma^2 + \sum_{j\neq m,n}^{2N} t_j p_j |h_{nj}|^2},
    \label{eq:flex_sinr}
\end{equation}
where $p_m$ denotes the transmit power of the $m$th node, $h_{n,m} \in \mathbb{C}$ denotes the wireless channel from $U_m$ to $U_n$, and $\sigma^2$ denotes the complex Gaussian noise power. The term $\sum_{j\neq m,n}^{2N} t_j p_j |h_{nj}|^2$ represents the total interference from $(N-1)$ transmitter nodes. The binary variable $t_m \in \{0, 1\}$ indicates whether $U_m$ is a transmitter or a receiver with $t_m=1$ and $t_m=0$, respectively. This variable ensures that transmitter nodes have zero received SINR and receiver node channels are excluded from the interference expression. Additionally, to analyze the more challenging generalized scenario, we assume non-reciprocal channels, i.e., $h_{nm} \neq h_{mn}$.

Coordinated eavesdroppers attempt to intercept FlexD communications by treating individual links as desired signals and considering others as interference. This process can be parallelized to decode all the communications from FlexD user pairs.
Then, the received SINR of FlexD user $m$ at the coordinated eavesdroppers' end, assuming MMSE receivers, is given by
\vspace{-1em}
\begin{equation}
    \gamma_m^E = t_m p_m \bm{g}_m^\dag \left( \sigma^2\bm{I}_K + \sum_{j \neq m,n}^{2N} t_j p_j \bm{g}_j \bm{g}_j^\dag \right)^{-1} \bm{g}_m
    \label{eq:eve_sinr}
\end{equation}
where $\bm{g}_m \in \mathbb{C}^{K \times 1}$ and $\bm{I}_K$ denotes the channel vector from the $m$th FlexD user to $K$ eavesdroppers and the identity matrix, respectively.

Using \eqref{eq:flex_sinr} and \eqref{eq:eve_sinr}, the instantaneous secrecy rate achieved by the communication from $m$th user to $n$th user is expressed as \cite{secrecy_2006}
\vspace{-0.5em}
\begin{equation}
    C_{mn} = \left[ \ln \left( 1 + \gamma_n^F \right) - \ln \left( 1 + \gamma_m^E \right) \right]^{+}.
    \label{eq:secrecy_rate_single}
\end{equation}
This expresses the achievable secrecy rate by selecting communication from $m$ to $n$, with a similar expression for $n$ to $m$. One expression evaluates to zero based on the communication direction.
The instantaneous secrecy rate expression can be evaluated for all user pairs and communication directions to obtain the sum secrecy rate of the FlexD network. This rate is assured over a fixed channel state in a given resource block. In the next subsection, we formulate the objective function to maximize the sum secrecy rate for all user pairs in the network.

\subsection{Optimization Problem}

The optimization goal is to maximize the sum secrecy rate of the FlexD network, as shown in \eqref{eq:secrecy_rate_single}. This entails optimizing power and transmission direction for all user pairs. The objective is mathematically represented as
\begin{subequations}\label{eq:optimization_objective}
\begin{align}
    \max_{\bm{p}, \bm{t}} &\sum_{n=1}^{2N} C_{mn}, \\
    \text { s.t. } \quad &0 \leq p_{n} \leq P_{\max}, \quad \forall n, \label{c_1} \\
    &m = 2(n\bmod2) + n - 1, \quad \forall n, \label{c_2} \\
    &t_m = 1 - t_n, \quad \forall n, \label{c_3}  \\
    &t_n \in \{0, 1\}, \quad \forall n, \label{c_4} 
\end{align}
\end{subequations}
where $P_{\max}$ is the maximum transmit power of a FlexD node. 

Notably, Addressing this challenge involves grappling with various complexities: i) the problem's inherent nature is marked by non-convexity, non-differentiability, and a combinatorial structure, making it inherently intricate; ii) the computational burden is heightened due to the inclusion of high-dimensional vector spaces; iii) the absence of readily available optimal solutions hampers seamless integration with supervised learning methods; and iv) real-world scenarios present the added hurdle of lacking eavesdroppers' channel data, compelling legitimate users to rely on location/distance data for estimating channel conditions. 
Addressing the  challenges of this problem demands innovative solutions. In Section~\ref{sec:fullCSI}, we tackle complexities i), ii), and iii) by using classical methods and GNN approaches. Subsequently, in Section~\ref{sec:NoFullCSI}, we extend our GNN approach to address complexity iv) related to the unavailability of full eavesdroppers' channel knowledge. 

\section{Sum Secrecy Rate Maximization with Full CSI}\label{sec:fullCSI}

Addressing non-convexity and the combinatorial nature, both classical and GNN solutions employ iterative algorithms. This involves transforming Problem \eqref{eq:optimization_objective} into a differentiable function by relaxing the non-negative constraint. As an optimal solution for Problem \eqref{eq:optimization_objective} is unavailable, the proposed GNN model employs unsupervised learning for training. Notably, the GNN solution exhibits quadratic time complexity, ensuring computational efficiency and suitability for implementation on low-level hardware.

\subsection{Classical Approach}

We introduce an iterative algorithm designed to address a relaxed version of the optimization objective presented in \eqref{eq:optimization_objective}. Due to the absence of differentiability, we eliminate the non-negative operator in \eqref{eq:secrecy_rate_single}. Consequently, the optimization problem can be projected into a higher-dimensional space using the MMSE-SINR equivalence. The resulting relaxed problem is then solved iteratively for $\bm{p}$ and $\bm{t}$, employing algorithms proposed in \cite{xu2019weighted} and \cite{Dayarathna2021}, respectively.

This block-coordinate descent mechanism is repeated until the sum secrecy rate converges to a user-defined threshold. However, it is worth noting that this algorithm exhibits diminished performance due to the challenges posed by the high-dimensional optimization landscape. Specifically, in scenarios with an increased number of users in the network, the algorithm optimizing $\bm{t}$ becomes predominant, resulting in a time complexity of $\mathcal{O}(N^4)$ under the assumption of linear time complexity for determining the sum secrecy rate. Conversely, in situations with more eavesdroppers in the network, the algorithm optimizing $\bm{p}$ takes precedence, leading to a time complexity of $\mathcal{O}(NK^3)$. This elevated time complexity can pose a computational bottleneck, especially in dense real-world scenarios.

\subsection{Graph Neural Network (GNN) Approach}
\begin{figure}[!t]
    \centering
    \tikzset{every picture/.style={line width=0.75pt}} %

\begin{tikzpicture}[x=0.75pt,y=0.75pt,yscale=-1,xscale=1]

\draw  [color={rgb, 255:red, 102; green, 166; blue, 30 }  ,draw opacity=1 ][line width=3.25]  (300,61.5) .. controls (300,46.86) and (311.86,35) .. (326.5,35) .. controls (341.14,35) and (353,46.86) .. (353,61.5) .. controls (353,76.14) and (341.14,88) .. (326.5,88) .. controls (311.86,88) and (300,76.14) .. (300,61.5) -- cycle ;
\draw  [color={rgb, 255:red, 102; green, 166; blue, 30 }  ,draw opacity=1 ][line width=3.25]  (154,88.5) .. controls (154,73.86) and (165.86,62) .. (180.5,62) .. controls (195.14,62) and (207,73.86) .. (207,88.5) .. controls (207,103.14) and (195.14,115) .. (180.5,115) .. controls (165.86,115) and (154,103.14) .. (154,88.5) -- cycle ;
\draw  [color={rgb, 255:red, 102; green, 166; blue, 30 }  ,draw opacity=1 ][line width=3.25]  (323,176.5) .. controls (323,161.86) and (334.86,150) .. (349.5,150) .. controls (364.14,150) and (376,161.86) .. (376,176.5) .. controls (376,191.14) and (364.14,203) .. (349.5,203) .. controls (334.86,203) and (323,191.14) .. (323,176.5) -- cycle ;
\draw [color={rgb, 255:red, 0; green, 0; blue, 0 }  ,draw opacity=0.5 ][line width=1.5]    (207,102) -- (323,163) ;
\draw [color={rgb, 255:red, 0; green, 0; blue, 0 }  ,draw opacity=0.5 ][line width=1.5]    (210,83) -- (297,67) ;
\draw [color={rgb, 255:red, 0; green, 0; blue, 0 }  ,draw opacity=0.5 ][line width=1.5]    (333,91) -- (345,147) ;
\draw  [color={rgb, 255:red, 102; green, 166; blue, 30 }  ,draw opacity=0.4 ][line width=3.25]  (172,162.5) .. controls (172,147.86) and (183.86,136) .. (198.5,136) .. controls (213.14,136) and (225,147.86) .. (225,162.5) .. controls (225,177.14) and (213.14,189) .. (198.5,189) .. controls (183.86,189) and (172,177.14) .. (172,162.5) -- cycle ;
\draw  [color={rgb, 255:red, 102; green, 166; blue, 30 }  ,draw opacity=0.4 ][line width=3.25]  (400,139.5) .. controls (400,124.31) and (412.31,112) .. (427.5,112) .. controls (442.69,112) and (455,124.31) .. (455,139.5) .. controls (455,154.69) and (442.69,167) .. (427.5,167) .. controls (412.31,167) and (400,154.69) .. (400,139.5) -- cycle ;
\draw [color={rgb, 255:red, 0; green, 0; blue, 0 }  ,draw opacity=0.2 ]   (187,118) -- (191,134) ;
\draw [color={rgb, 255:red, 0; green, 0; blue, 0 }  ,draw opacity=0.2 ]   (297,67) -- (204,133) ;
\draw [color={rgb, 255:red, 0; green, 0; blue, 0 }  ,draw opacity=0.2 ]   (319,173) -- (228,167) ;
\draw [color={rgb, 255:red, 0; green, 0; blue, 0 }  ,draw opacity=0.2 ]   (378,167) -- (402,157) ;
\draw [color={rgb, 255:red, 0; green, 0; blue, 0 }  ,draw opacity=0.2 ]   (404,120) -- (351,79) ;
\draw [color={rgb, 255:red, 0; green, 0; blue, 0 }  ,draw opacity=0.2 ]   (397,135) -- (207,102) ;
\draw [color={rgb, 255:red, 0; green, 0; blue, 0 }  ,draw opacity=0.2 ]   (397,135) -- (227,157) ;

\draw (308.4,50.7) node [anchor=north west][inner sep=0.75pt]   [align=left] {$\bm{x}_{\mathcal{G}}^{\ell } (n)$};
\draw (162.4,78.7) node [anchor=north west][inner sep=0.75pt]   [align=left] {$\bm{x}_{\mathcal{G}}^{\ell } (1)$};
\draw (329.4,165.7) node [anchor=north west][inner sep=0.75pt]   [align=left] {$\bm{x}_{\mathcal{G}}^{\ell } (N)$};
\draw (232.54,58.27) node [anchor=north west][inner sep=0.75pt]  [font=\small,rotate=-349.78]  {$\bm{e}_{\mathcal{G}}^{\ell } (1,n)$};
\draw (250.76,104.78) node [anchor=north west][inner sep=0.75pt]  [font=\small,rotate=-29.33]  {$\bm{e}_{\mathcal{G}}^{\ell } (1,N)$};
\draw (351,87.35) node [anchor=north west][inner sep=0.75pt]  [font=\small,rotate=-77.02]  {$\bm{e}_{\mathcal{G}}^{\ell } (n,N)$};
\draw (180.4,152.7) node [anchor=north west][inner sep=0.75pt]  [color={rgb, 255:red, 0; green, 0; blue, 0 }  ,opacity=0.4 ] [align=left] {$\bm{x}_{\mathcal{G}}^{\ell } (2)$};
\draw (402.4,130.7) node [anchor=north west][inner sep=0.75pt]  [font=\scriptsize,color={rgb, 255:red, 0; green, 0; blue, 0 }  ,opacity=0.4 ] [align=left] {$\bm{x}_{\mathcal{G}}^{\ell } (N-1)$};
\draw (160,45) node [anchor=north west][inner sep=0.75pt]  [font=\scriptsize] [align=left] {Node $1$};
\draw (361,51) node [anchor=north west][inner sep=0.75pt]  [font=\scriptsize] [align=left] {Node $n$ };
\draw (399,96) node [anchor=north west][inner sep=0.75pt]  [font=\scriptsize,color={rgb, 255:red, 0; green, 0; blue, 0 }  ,opacity=0.4 ] [align=left] {Node $N-1$ };
\draw (177,193) node [anchor=north west][inner sep=0.75pt]  [font=\scriptsize,color={rgb, 255:red, 0; green, 0; blue, 0 }  ,opacity=0.4 ] [align=left] {Node $2$};
\draw (383,181) node [anchor=north west][inner sep=0.75pt]  [font=\scriptsize,color={rgb, 255:red, 0; green, 0; blue, 0 }  ,opacity=1 ] [align=left] {Node $N$};

\end{tikzpicture}
    \caption{The proposed graph representation of the FlexD network.}
    \label{fig:graph_rep}
    \vspace{-1.2em}
\end{figure}
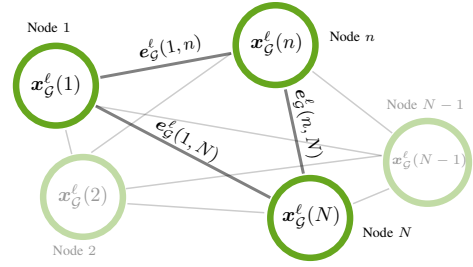

The classical approach to \eqref{eq:optimization_objective} is computationally complex and suboptimal due to coordinate descent's susceptibility to converge to local optima. we propose a novel GNN-based unsupervised learning strategy for maximizing sum secrecy rate in the FlexD network. Unlike traditional methods, our approach concurrently optimizes $\bm{t}$ and $\bm{p}$, reducing computational complexity to $\mathcal{O}(N^2)$ with our proposed graph structure as input.

\subsubsection{GNNs}

Learning from graph-structured data presents a unique challenge, requiring the preservation of the underlying graph structure. GNNs effectively address this challenge, consistently yielding promising results in diverse applications \cite{guan2021customized, wang2020learning, zhang2021indoor}. GNNs leverage neighborhood, node, and edge information to iteratively learn node representations within a graph. Most GNN architectures employ information aggregation and combination strategies across successive iterations, commonly known as graph layers. The aggregation and combination steps in the $\ell$th layer can be formally expressed as:
\begin{equation}
    \bm{x}_{\mathcal{G}}^{\ell}(n) = f\left( \bm{x}_{\mathcal{G}}^{\ell-1}(n), \{\!\!\{\bm{x}_{\mathcal{G}}^{\ell-1}(j):j \in \mathcal{N}(n)\}\!\!\}  \right),
\end{equation}
where the function $f$ encapsulates the specific neural network architecture tailored to the given use case. The notation $\mathcal{N}(v)$ denotes the neighbors of node $v$. This iterative process allows GNNs to effectively capture and utilize the structural information inherent in graph data, showcasing their adaptability and efficacy in complex learning tasks.

\subsubsection{Proposed Graph Representation of the FlexD Network}
Our proposed network model manifests as an undirected graph $\mathcal{G} = (\mathcal{V}, \mathcal{E})$, where $\mathcal{V}$ and $\mathcal{E}$ denote the vertices and edges of the graph, respectively. Each node in the graph corresponds to a user pair, resulting in a total of $N$ nodes. The node features encapsulate two non-reciprocal channels between the $m$th  and $n$th users, denoted as $\{\!\!\{\bm{x}_{\mathcal{G}}(n) = h_{mn} |\!| h_{nm}:n \in \mathcal{V}\}\!\!\}$. The interference from user pairs to other user pairs is represented by the edges of the graph, encapsulated as edge features $\{\!\!\{\bm{e}_{\mathcal{G}}(n, j) = h_{nj} |\!| h_{jn} |\!| h_{mj} |\!| h_{jm}: \{n,j\} \in \mathcal{E} \}\!\!\}$. The proposed graph representation is depicted in Fig.~\ref{fig:graph_rep}.

It is noteworthy that eavesdroppers, being passive listeners, are not represented as vertices in the graph. Channel information from users to eavesdroppers is integrated through linear projections to the node embedding within the graph neural architecture.

\subsubsection{Proposed GNN Architecture}
In our prior research on sum rate maximization of FlexD networks \cite{flex_net_2022}, we employed an edge-level task to determine communication direction and a node-level task to identify the power vector of a graph. This work introduces a more computationally and memory-efficient graph structure, extracting both direction and power vectors through a node-level task. Notably, this approach eliminates the need for two distinct aggregation schemes.

The proposed graph structure, depicted in Fig.~\ref{fig:graph_rep}, leverages node features derived from paired channel information and linear projections of channel data to eavesdroppers. The linear projection is mathematically expressed as:
\vspace{-0.2em}
\begin{equation}\label{eq:projection_with_channel_data}
    \bm{y}(n) = \bm{W}_p \left(\bm{g}_m |\!| \bm{g}_n \right),
\end{equation}
\vspace{-0.3em}where $\bm{W}_p \in \mathbb{C}^{c \times (|\bm{g}_m|+|\bm{g}_n|)}$ represents a learnable projection matrix, with $c=8$ in our experiments. This projection serves to reduce the dimensionality of the channel information vector, enhancing learning performance and computational efficiency. The node features of the input graph can be expressed as:
\vspace{-0.2em}
\begin{equation}
    \bm{x}_{\mathcal{G}}^0(n) = h_{mn} |\!| h_{nm} |\!| \bm{y}(n).
\end{equation}
\vspace{-0.2em}Here, the superscript $0$ denotes the input graph. For a given edge, the input edge features encompass the {\it eight} different \textit{potential} interference channels:
\vspace{-0.2em}
\begin{equation}
    \bm{e}_{\mathcal{G}}^0(n, j) = \big\Vert_{s \in S} h_s : S=\{ nj, jn, mj, jm, ni, in, mi, im\}
\end{equation}
\vspace{-0.2em}where $n$ and $j$ represent the connecting nodes of the edge, $m~=~2(n\bmod2) + n - 1$ and $i\!=\!2(j\bmod2) + j - 1$.
The processed input graph is then fed into the GNN to learn representations of node and edge features. The aggregation function is represented as two nested functions. The aggregation function for the $\ell$th layer can be written as:
\vspace{-0.3em}
\begin{align}
    \bm{x}_{\mathcal{G}}^{\ell}(n) &= \varphi\left( \sum_{j\in\mathcal{N}(n)} \bm{W}_s^{\ell-1} \bm{x}_{\mathcal{G}}^{\ell-1}(n) \right.\nonumber \\
    & \qquad \qquad +\bm{W}_i^{\ell-1} \bm{x}_{\mathcal{G}}^{\ell-1}(j) + \bm{W}_e^{\ell-1} \bm{e}_{\mathcal{G}}^{\ell-1}(n,j)\Biggl),
\end{align}
\vspace{-0.2em}where $\varphi$ is a chosen nonlinearity function\footnote{ $\operatorname{GELU}$ is used in our experiments. Unlike $\operatorname{RELU}$, this activation function has a well-defined gradient in the negative region, providing a solution to the dying neurons in the negative region.}, and $\bm{W}_s^{\ell-1} \in \mathbb{C}^{d_\ell \times |\bm{x}_{\mathcal{G}}^{\ell-1}|}$, $\bm{W}_i^{\ell-1} \in \mathbb{C}^{d_\ell \times |\bm{x}_{\mathcal{G}}^{\ell-1}|}$, and $\bm{W}_e^{\ell-1}\in \mathbb{C}^{d_\ell \times |\bm{e}_{\mathcal{G}}^{\ell-1}|}$ are weight matrices for node embedding, neighbor node embedding, and edge embedding, respectively. $d_\ell$ represents the cardinality of node embeddings of the $\ell$th layer.
This aggregation is repeated for $\mathcal{L}$ layers, simulating an iterative message-passing mechanism. The $\mathcal{L}$th layer contains node embeddings that capture the structural information of the entire input graph, enabling node-level decisions regarding power value and communication direction. The transmit power of the $n$th node pair is obtained using:
\begin{equation}\label{eq:power_out}
    p(n) = P_{\max}.\phi\left[f_p\left(\bm{x}_{\mathcal{G}}^{\mathcal{L}}(n)\right)\right],
\end{equation}
where $p(n)\in\mathbb{R}$ is the power value of the $n$th user pair, $\phi$ denotes the sigmoid activation function, and $f_p$ is approximated by a multi-layer perception (MLP). The sigmoid activation function constrains power values to the $(0, 1)$ range, scaled to $(0, P_{\max})$ by multiplying by $P_{\max}$. The direction value of the user pair $n$ is found using:
\begin{equation}\label{eq:dir_out}
   \bm{t}(n) = \varsigma\left[f_t\left(\bm{x}_{\mathcal{G}}^{\mathcal{L}}(n)\right)\right], 
\end{equation}
where $\bm{t}(n)\in\mathbb{R}^{2 \times 1}$ is the direction vector of the $n$th user pair, $\varsigma$ denotes the softmax function, and $f_t$ is approximated by an MLP.

\subsubsection{Loss Function in the Unsupervised Learning Setting}
In optimizing the trainable weights within our GNN model, we employ an improved variant of the stochastic gradient descent (SGD) algorithm known as \textit{AdamW}. Despite the non-differentiable nature of problem \eqref{eq:optimization_objective} due to the non-negative operator, sub-derivatives in conjunction with SGD can be used to train the GNN model. 

However, our experiments reveal that training performance improves when utilizing a relaxed version of the loss function. The loss function, derived from the relaxed version of problem \eqref{eq:optimization_objective}, is expressed as:
\vspace{-0.3em}
\begin{align}
    \min_{\bm{p}, \bm{t}} &-\sum_{n=1}^{2N}\tilde{C}_{mn}, \\
    \text { s.t. }\quad & \tilde{C}_{mn} = \ln \left( 1 + \gamma_n^F \right) - \ln \left( 1 + \gamma_m^E \right),  \nonumber \\
    &\eqref{c_1}, \eqref{c_2}, \eqref{c_3}, \eqref{c_4}. \nonumber
\end{align}
Here, the negative value of the relaxed sum secrecy rate is employed to transform the problem into a standard neural network minimization task.

\section{Sum Secrecy Rate Maximization without Eavesdroppers' CSI}\label{sec:NoFullCSI} 
In real-world scenarios, legitimate users often lack perfect channel state information (CSI) of eavesdroppers' channels. Instead, they rely on statistical information derived from factors like location and distance to eavesdroppers. This reliance on statistical information poses challenges when maximizing the sum secrecy rate, particularly with algorithms designed for perfect CSI scenarios. Statistical models effectively capture wireless multipath fading channels in practical situations, accounting for large-scale and small-scale fading effects. Distance information, acting as a proxy, becomes valuable in situations where channel information is unavailable.
\begin{thm}\label{thm:universal_prob}
Given the distance matrix $\bm{D}\in\mathbb{R}^{2N\times K}$ to eavesdroppers, there exists an MLP-based approximation $f_{\operatorname{MLP}}~:~\bm{D}~\mapsto~\mathbb{C}^{2N \times K}$ such that, for any accuracy threshold $\varepsilon > 0$, the channel matrix $\bm{G}\in\mathbb{C}^{2N \times K}$ can be approximated as $\hat{\bm{G}}$ with $|\bm{G} - \hat{\bm{G}}| \leq \varepsilon$.
\end{thm}
\begin{IEEEproof} Based on Theorem 2.1 \cite{lu2020universal}, given the input and output distributions $P(\bm{D})$ and $P(\bm{G})$, respectively, under specific integrability conditions, the probability measure $P(\bm{G})$ can be approximated to any given error $\varepsilon$ by transforming $P(\bm{D})$ using the gradient of a potential function. This transformation can be parameterized by an MLP.
\end{IEEEproof}
Theorem \ref{thm:universal_prob} states that MLPs can effectively map one distribution to another up to a certain precision. Leveraging this property, we can approximate the channel distribution of eavesdroppers by utilizing the distance metric of eavesdroppers. In the proposed graph representation, the distance to each eavesdropper replaces channel information when computing the linear projection \eqref{eq:projection_with_channel_data}. This mathematical representation can be expressed as:
\vspace{-0.3em}
\begin{equation}\label{eq:projection_with_distance}
    \bm{y}(n) = \tilde{\bm{W}}_p \left(\bm{d}_m |\!| \bm{d}_n \right)
\end{equation}
\vspace{-0.2em}where $\tilde{\bm{W}}_p \in \mathbb{C}^{c \times (|\bm{d}_m|+|\bm{d}_n|)}$ denotes a trainable weight matrix. $\bm{d}_m$ and $\bm{d}_n$ represent distance vectors to eavesdroppers from the $m$th and $n$th users, respectively.

\section{Numerical Results}
We compare proposed approaches through simulations, evaluating performance and time complexity. We also explore HD and FlexD communications performance, along with the GNN approach with and without eavesdroppers' full CSI.

\vspace{-0.2em}
\subsection{Simulation Setup}
We consider a network consisting of $N$ user pairs and $K$ eavesdroppers which are spanned over $1 \times 1\;\text{km}^2$ area. User pairs and the eavesdroppers are Poisson disk distributed over the given area. Users are paired randomly to form user pairs in the network. Channel realizations are generated by assuming Rayleigh fading with free space path loss at $1\; \text{GHz}$ frequency and $8\;\text{dB}$ log-normal shadowing. Peak power of $30\;\text{dBm}$ is assumed along with a noise level of $-100\;\text{dBm}$. 

Further, simulation results are presented for two additional baseline strategies: HD, the classical approach without direction selection, and \textit{Max Power}, where the direction with the strongest channel of the desired link is selected, and $P_{\max}$ is used as the transmit power.
All the algorithms are implemented in Python language and tested on CPU for time comparisons.
The GNN model consists of 3 layers and the MLPs used to realize \eqref{eq:power_out} and \eqref{eq:dir_out} are kept consistent throughout the experiments. GNN models are trained with 10000 network realizations with a batch size of 128 samples. The \textit{AdamW} optimizer is used with the learning rate of $0.002$. 

\vspace{-0.2em}
\subsection{Performance Comparison}
\begin{figure}[!t]
    \centering
    \includegraphics[width=0.8\linewidth]{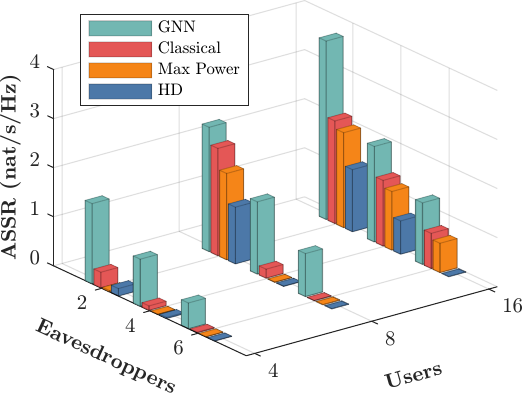}
    \caption{Average sum secrecy rate (ASSR) performance comparison of GNN model, classical algorithm, \textit{Max Power}, and HD communication.}
    \label{fig:performance}
    \vspace{-1.2em}
\end{figure}

Fig.~\ref{fig:performance} presents a  comparison of the average sum secrecy rate (ASSR) performance across varying user and eavesdropper counts. The inherent limitations imposed by the coordinated nature of eavesdroppers are evident in the observed spectral efficiency, thereby providing valuable insights into the network's capabilities under challenging conditions.

Notably, the GNN approach outperforms the classical approach, \textit{Max Power}, and HD communications by a substantial margin. For instance, with $4$ users and $2$ eavesdroppers, the GNN achieves a performance gain of $5$-fold and $10$-fold compared to classical and HD scenarios, respectively. However, this performance discrepancy diminishes with an increasing number of users and decreasing number of eavesdroppers due to the diversity introduced by more desired channels and the reduction in leakage links. A noteworthy observation is that the GNN model maintains a constant learnable parameter count, excluding the linear projection matrix, throughout the experiments. This highlights the scalability of the GNN, demonstrating minimal increases in learnable parameters and subsequently reducing computational costs, particularly in larger networks.

Across all scenarios, the classical algorithm with direction selection consistently outperforms the HD communications. This substantiates the claim that direction selection in FlexD networks significantly contributes to performance gains. For instance, in a $16$-user network with $2$ and $4$ eavesdroppers, the performance gain from FlexD or direction selection is $1.6$ times and $2$ times, respectively. Additionally, the superior performance of the \textit{Max Power} strategy with a better direction selection criterion validates this claim. 

\begin{figure*}[!t]
\centering
\subfloat[]{\includegraphics[width=0.31\linewidth]{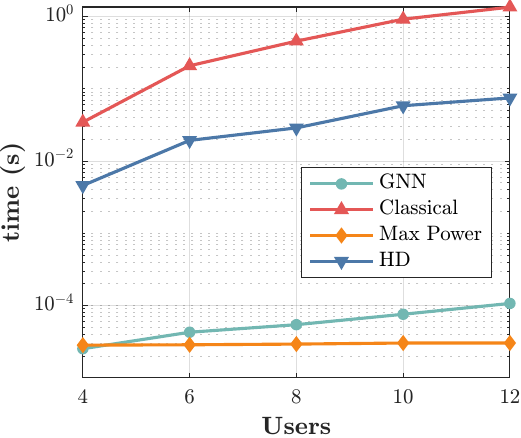}
\label{fig:time_users}}
\hfil
\subfloat[]{\includegraphics[width=0.31\linewidth]{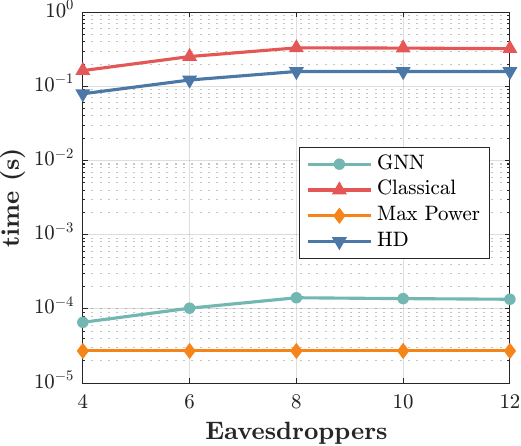}
\label{fig:time_eaves}}
\hfil
\subfloat[]{\includegraphics[width=0.31\linewidth]{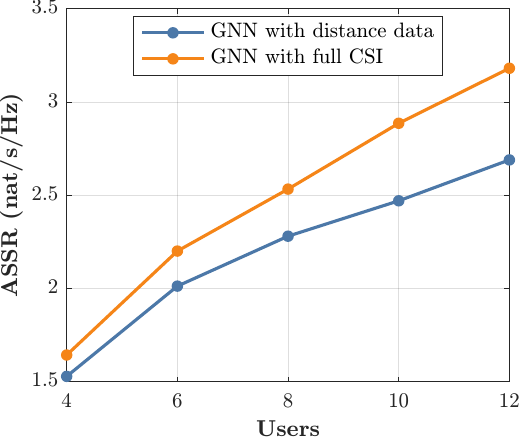}
\label{fig:location_perf}}
\caption{(a) Average running time of algorithms with $K=2$ eavesdroppers and a variable number of users. (b) Average running time of algorithms with 2 users and a variable number of eavesdroppers. (c) Average sum secrecy rate performance comparison of GNN models with and without eavesdropper channel knowledge when $K=2$ eavesdroppers are present in the network.}
\vspace{-1em}
\end{figure*}
\vspace{-0.2em}
\subsection{Time Complexity}
Fig.\ref{fig:time_users} and Fig.\ref{fig:time_eaves} show heuristic time complexity comparisons of algorithms. Each plot indicates single CPU time for algorithm execution, revealing insights into underlying mathematical operation time complexity.
\textit{Max Power} approach is computationally efficient than the GNN model but suffers from low performance due to the lack of proper power allocation. The GNN model consistently outperforms classical and HD communications with varying user and eavesdropper counts. This advantage becomes more pronounced with an increasing number of eavesdroppers while maintaining a fixed number of users. In Fig.\ref{fig:time_eaves} with $12$ eavesdroppers, the time gap is on the order of $10^{-3}$, while in Fig.\ref{fig:time_users} with $12$ users, it is on the order of $10^{-4}$. With an increasing number of eavesdroppers, the GNN approach maintains low time complexity growth due to dimensionality reduction in linear projection \eqref{eq:projection_with_channel_data}. As a result, the number of neurons needed for eavesdropper information processing remains nearly constant despite the growing number of eavesdroppers.
\vspace{-0.2em}
\subsection{Location Aided Secrecy Performance of the GNN Approach}

In Fig.~\ref{fig:location_perf}, we evaluate two GNN models: one using distance knowledge from users to eavesdroppers and the other incorporating full CSI. Distance knowledge proves effective as a proxy for channel information, yielding substantial performance. However, with an increasing number of users, a growing performance gap between the two approaches appears. This implies neural networks may have limitations in fully extracting diversity information solely from distance knowledge.
\section{Conclusion}
This paper addressed FlexD-based PLS challenges in multi-user networks by introducing a novel system model for multiple users and coordinated eavesdroppers. We formulated and optimized the sum secrecy rate using both a heuristic classical algorithm and a GNN-based strategy. 
Our contributions include an innovative system model, an efficient graph representation for FlexD network, and a GNN-based optimization outperforming the classical method in sum secrecy performance and time complexity. Notably, the proposed GNN achieves a lower complexity of $\mathcal{O}(N^2)$ compared to the classical method's worst-case complexity of $\mathcal{O}(N^4)$. 
Leveraging neural networks, we used distance data as a proxy for eavesdroppers' channel information, achieving reasonable performance. Future work will focus on user-fairness and improving partial CSI estimation through learning. 
\bibliographystyle{IEEEtran}
\bibliography{IEEEabrv,refs}

\end{document}